\documentclass[twocolumn,twoside,slac_two]{revtex4}
\usepackage{graphicx}
\usepackage{fancyhdr}
\begin{document}

\title{Presymmetry beyond the Standard Model}

\author{Ernesto A. Matute}
\affiliation{Departamento de F\'{\i}sica, Universidad de Santiago
de Chile, Casilla 307, Santiago, Chile}

\begin{abstract}
We go beyond the Standard Model guided by presymmetry, the
discrete electroweak quark-lepton symmetry hidden by topological
effects which explain quark fractional charges as in condense
matter physics. Partners of the particles of the Standard Model
and the discrete symmetry associated with this partnership appear
as manifestations of a residual presymmetry and its extension from
matter to forces. This duplication of the spectrum of the Standard
Model keeps spin and comes nondegenerated about the TeV scale.
\end{abstract}

\maketitle

\thispagestyle{fancy}

\section{Introduction}
The notion of presymmetry was introduced by
Ekstein~\cite{Ekstein1,Ekstein2} in the sixties to deal with the
survival of some results of space-time symmetry when this is
broken by an external field. It is a pre-dynamical symmetry which
becomes only partially broken by the dynamics with a residual
presymmetry. We have used the same term in the Standard Model (SM)
of elementary particle physics and done analogies to the idea of
Ekstein~\cite{EAM1,EAM2}.

In the SM, presymmetry becomes a symmetry which extends the
quark-lepton symmetry from weak to electromagnetic interactions.
It is a hidden charge symmetry, so hidden that it can be
eliminated by using the Occam's razor principle which states that
entities should not be multiplied unnecessary, unless that some
results of the charge symmetry survive in the new physics beyond
the SM~\cite{EAM2}.

This work has been organized as follows. We start by presenting
the quark-lepton charge symmetry that has motivated the research
(Section 2). Next, we state hypotheses to account for this charge
symmetry (Section 3). We describe the approach, which adds to the
SM the new hidden states of prequarks and preleptons, and the
associated presymmetry (Section 4). The problem of gauge anomaly
is addressed (Section 5). Motivations to go beyond the SM with
presymmetry are given, emphasizing a duplication of the SM particles
to have a residual presymmetry in the sense of Ekstein (Section 6).
And we finish with some conclusions based on results (Section 7).

\section{Quark-Lepton Charge Symmetry \label{QLsym}}
\begin{figure}
\centering
\includegraphics[width=80mm]{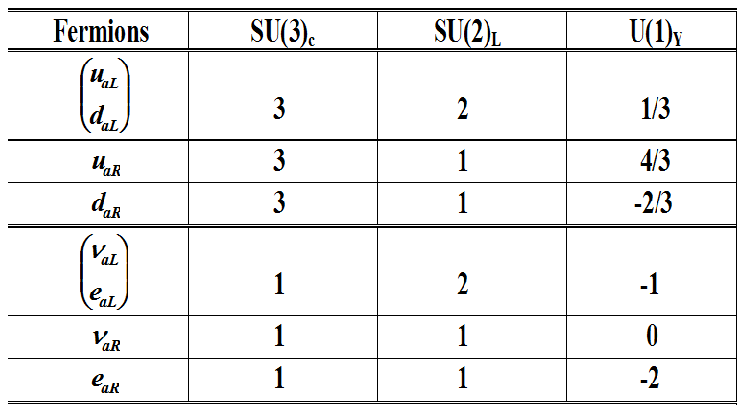}
\caption{Quark and lepton assignments to representations of gauge
groups of the SM, where ${\it a}$ denotes the generation index.}
\label{SM}
\end{figure}

At the level of the SM, presented in Figure~\ref{SM}, quarks and
leptons are quite different in the strong sector: quarks come in
triplets while leptons do in singlets of the color group. However,
they have similar properties in the weak sector, a fact known as
quark-lepton symmetry. Regarding hypercharge, values appear very
different and no charge symmetry is readily seen. But, a closer
inspection shows that there is something underlying these values,
a hidden charge symmetry which extends the weak connection.

\begin{figure}
\centering
\includegraphics[width=82mm]{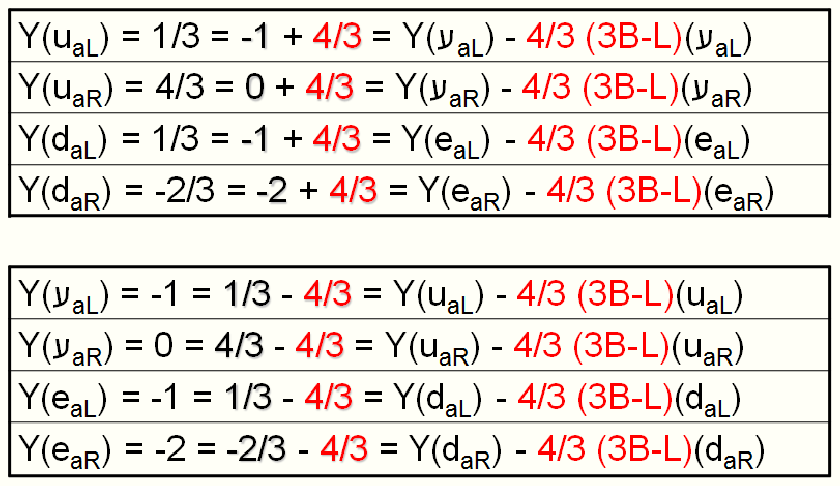}
\caption{Quark-lepton charge symmetry.} \label{table1}
\end{figure}

The one-to-one correspondence between quark and lepton
hypercharges is shown in Figure~\ref{table1}. The fractional
hypercharge of a quark is that of its lepton weak partner plus a
global fractional value which is independent of flavor and
handness. Actually, it is a fraction of the lepton number of its
partner. In a similar way, the entire hypercharge of a lepton is
that of its quark partner plus a global fractional value which is
a fraction of the baryon number times the number of colors.
Although this global fractional part depends upon the hypercharge
normalization, the quark-lepton charge symmetry is still present.

The crucial question is whether these charge relations are real or
accidental. Our hypothesis is that this quark-lepton charge
symmetry is real and that the global fractional piece of charge
has a topological character independently of the normalization
used for hypercharge~\cite{EAM1}.

\section{Statement of Principles}
Formally, our assumptions are described in these two
principles~\cite{EAM1}:
\\[10pt]
$\bullet$ Principle of electroweak quark-lepton symmetry:
\emph{There exists a hidden discrete $Z_{2}$ symmetry in the
electroweak interactions of quarks and leptons}.
\\[10pt]
$\bullet$ Principle of weak topological-charge confinement:
\emph{Observable particles have no weak topological charge}.
\\[10pt]
\indent The principle of weak topological-charge confinement is
secondary to that of gauge confinement in the sense that
electroweak forces by themselves cannot lead to actual confinement
of topologically nontrivial particles.

We distinguish the electroweak-symmetric topological quarks and
topological leptons of our approach from the topologically trivial
quarks and leptons of the SM. The assignments of topological quarks
to the gauge groups of the SM are as usual quarks in Figure~\ref{SM},
whereas those of topological leptons are defined below. Topological
quarks and topological leptons of fractional charge do have hidden
charge structures and nontrivial topology. They are not mass
eigenstates and have associated weak gauge fields with vacuum states
having nonzero topological charge. Topological quarks and topological
leptons have a topological bookkeeping $Z_{3}$ charge, with +1
assigned to all of them. The 3 of this modulo charge in topological
quarks and topological leptons is due to the number of colors and the
number of generations~\cite{EAM1}, respectively, though topological
leptons do not confine. When the bookkeeping charge is 3, the set has
no topological charge and trivial topology. Three topological quarks
are equivalent to three standard quarks.

\section{Topological Quarks and Leptons}
To describe the charge structure of topological quarks, we
introduce the new states of \emph{prequarks} with integer charge and
trivial topology, as in leptons. Similarly, we introduce
topological leptons or \emph{preleptons}, with fractional charge and
nontrivial topology, just as in topological quarks~\cite{EAM1}. We
denote prequarks and preleptons by $\hat{q}$ and $\hat{\ell}$,
respectively.

Charge dequantization in topological quarks and preleptons is
according to Figure~\ref{table1} and has the following form:
\begin{eqnarray}
Y(q) & = & \displaystyle Y(\hat{q}) - \frac{4}{3} \,
(B-3L)(\hat{q}) \, , \nonumber \\ & & \nonumber \\ Y(\ell) & = &
\displaystyle Y(\hat{\ell}) - \frac{4}{3} \, (B-3L)(\hat{\ell}) \,
. \label{Ycharges}
\end{eqnarray}

\noindent The lepton number is times 3 because prequarks are the
ones that now have integer charge and preleptons now have
fractional charge.

\begin{figure}
\centering
\includegraphics[width=80mm]{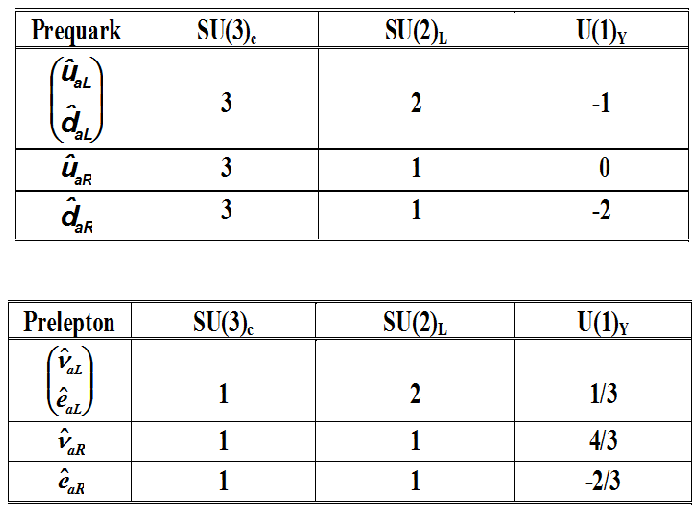}
\caption{Assignments of prequarks and preleptons to gauge groups
of the SM.} \label{slide2}
\end{figure}

Assignments of prequarks and preleptons to gauge groups of the SM
are shown in Figure~\ref{slide2} and charge connections with
leptons and topological quarks are given by the following
constraints consistent with relations in Figure~\ref{table1}:

\begin{eqnarray}
& & Y(\hat{q}) = Y(\ell) \, , \nonumber \\ \nonumber & & \\ & &
Y(\hat{\ell}) = Y(q) \, , \nonumber \\ & & \nonumber \\ & &
(B-3L)(\hat{q}) = (3B-L)(\ell) = -(3B-L)(q) \, , \nonumber \\ & &
\nonumber \\ & & (B-3L)(\hat{\ell}) = (3B-L)(q) = -(3B-L)(\ell) \,
. \label{Yconstraints}
\end{eqnarray}

\begin{figure}[t]
\centering
\includegraphics[width=80mm]{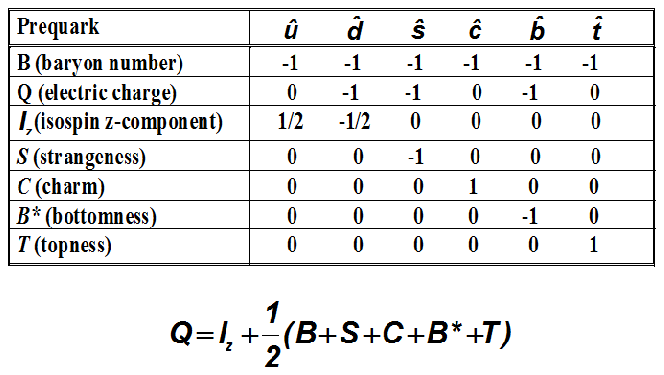}
\caption{Additive quantum numbers of prequarks.} \label{extra1}
\end{figure}

\begin{figure}[t]
\centering
\includegraphics[width=80mm]{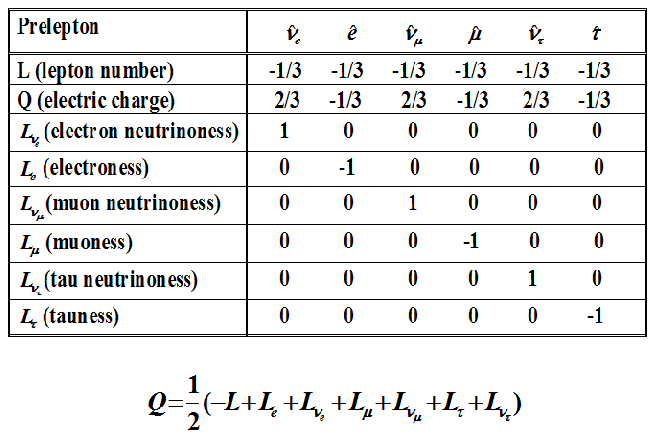}
\caption{Additive quantum numbers of preleptons.} \label{extra2}
\end{figure}

The additive quantum numbers of prequarks and preleptons are
listed in Figures~\ref{extra1} and \ref{extra2}, respectively. We
note that $B-L$ becomes a good quantum number between prequarks
and leptons, and between preleptons and topological
quarks~\cite{EAM1}.

\begin{figure}[h]
\centering
\includegraphics[width=70mm]{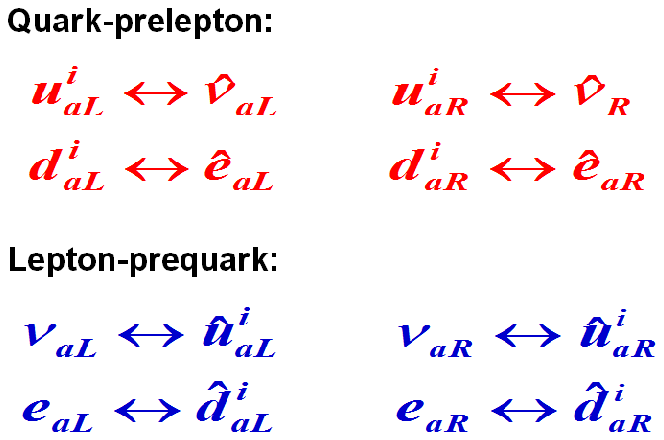}
\caption{Presymmetry between topological quarks and preleptons,
and between leptons and prequarks.} \label{slide1}
\end{figure}

This is a conservative model in the sense that usual quarks and
leptons are considered elementary with no charge structure. Charge
structure is in topological quarks and preleptons. There is
symmetry between topological quarks and preleptons, as between
prequarks and leptons.

Presymmetry is the statement of charge symmetry between
topological quarks and preleptons, and between leptons and
prequarks. There is invariance in the bare electroweak Lagrangian
under the flavor transformation given in Figure~\ref{slide1}, with
no change on gauge and Higgs fields.

\section{Gauge Anomaly Cancellation}
There is, however, a problem: the entire hypercharge of prequarks
and the fractional hypercharge of preleptons lead to gauge
anomalies. Here it is shown the anomaly in the scenario of
prequarks. It is found (see~\cite{EAM1}) that the
$\mbox{U(1)}_{Y}$ gauge current of prequarks and leptons

\begin{eqnarray}
\hat{J}^{\mu}_{Y} & = & \displaystyle \overline{\hat{q}}_{aL}
\gamma^{\mu} \frac{Y}{2} \hat{q}_{aL} + \overline{\hat{q}}_{aR}
\gamma^{\mu} \frac{Y}{2} \hat{q}_{aR} \nonumber \\
& & \displaystyle + \; \overline{\ell}_{aL} \gamma^{\mu}
\frac{Y}{2} \ell_{aL} + \overline{\ell}_{aR} \gamma^{\mu}
\frac{Y}{2} \ell_{aR} \, ,
\end{eqnarray}

\noindent exhibits the $\mbox{U(1)}_{Y} [\mbox{SU(2)}_{L}]^{2}$
and $[\mbox{U(1)}_{Y}]^{3}$ anomalies according to

\begin{equation}
\begin{array}{rcl}
\partial_{\mu} \hat{J}^{\mu}_{Y} & = &  \displaystyle -
\frac{g^{2}}{32 \pi^{2}} \left( \, \sum_{\hat{q}_{L} \ell_{L}}
\frac{Y}{2} \right) \mbox{tr} \; W_{\mu\nu} \tilde{W}^{\mu\nu} \\
& & \\ & & \displaystyle - \frac{{g'}^{2}}{48 \pi^{2}} \left( \,
\sum_{\hat{q}_{L} \ell_{L}} \frac{Y^3}{2^3} - \sum_{\hat{q}_{R}
\ell_{R}} \frac{Y^3}{2^3} \right) F_{\mu\nu} \tilde{F}^{\mu\nu} \,
,
\end{array}
\end{equation}
\\[3pt]
where $g$, $g'$ and $W_{\mu\nu}$, $F_{\mu\nu}$ are the
$\mbox{SU(2)}_{L}$, $\mbox{U(1)}_{Y}$ couplings and field
strengths, respectively, $\tilde{W}^{\mu\nu}$ and similarly
$\tilde{F}^{\mu\nu}$ is defined by

\begin{equation}
\tilde{W}_{\mu\nu} = \frac{1}{2} \; \epsilon_{\mu\nu\lambda\rho}
\; W^{\lambda\rho} \, ,
\end{equation}

\noindent and the hypercharge sums run over all the fermion
representations. The anomalies appear because these sums do not
vanish.

The anomaly can be written in terms of topological currents as
follows:

\begin{equation}
\partial_{\mu} \hat{J}^{\mu}_{Y} = - N_{\hat{q}} \; \partial_{\mu}
J^{\mu}_{T} \, ,
\end{equation}

\noindent where $N_{\hat{q}} = 12 N_{g}$ is the number of left-
and right-handed prequarks, $N_{g}$ denotes the number of
generations, and

\begin{eqnarray}
J^{\mu}_{T} & = & \frac{1}{4 N_{\hat{q}}} \, K^{\mu}
\sum_{\hat{q}_{L} \ell_{L}} Y \nonumber \\
& & + \frac{1}{16 N_{\hat{q}}} \, L^{\mu} \left( \,
\sum_{\hat{q}_{L} \ell_{L}} Y^{3} - \sum_{\hat{q}_{R}
\ell_{R}} Y^{3} \right) \nonumber \\
& = &- \, \frac{1}{6} \, K^{\mu} + \frac{1}{8} \, L^{\mu} \, .
\end{eqnarray}

\noindent The $K^{\mu}$ and $L^{\mu}$ are the well-known
topological currents or Chern--Simons classes related to the
$\mbox{SU(2)}_{L}$ and $\mbox{U(1)}_{Y}$ gauge groups,
respectively. They are given by

\begin{eqnarray}
\displaystyle K^{\mu} &=& \displaystyle \frac{g^{2}}{8 \pi^{2}} \,
\epsilon^{\mu\nu\lambda\rho} \; \mbox{tr} \!\! \left( W_{\nu}
\partial_{\lambda} W_{\rho} - \frac{2}{3} \, i g
W_{\nu} W_{\lambda} W_{\rho} \right) , \nonumber \\ & & \nonumber
\\ \displaystyle L^{\mu} &=& \displaystyle \frac{{g'}^{2}}{12
\pi^{2}} \, \epsilon^{\mu\nu\lambda\rho} A_{\nu}
\partial_{\lambda} A_{\rho} \, .
\end{eqnarray}

The counterterm needed for anomaly cancellation is
\begin{equation}
\Delta {\cal L} = g' N_{\hat{q}} \, J^{\mu}_{T} A_{\mu} \, .
\end{equation}

\noindent It leads to the anomaly-free, but gauge noninvariant
current

\begin{equation}
J^{\mu}_{Y} = \hat{J}^{\mu}_{Y} + N_{\hat{q}} \; J^{\mu}_{T} \, ,
\label{newJ}
\end{equation}

\noindent such that
\begin{equation}
\partial_{\mu} J^{\mu}_{Y} = 0.
\end{equation}

However, the charge of the new current is not conserved because of
topological charge. Topological charge resides in the topology of
weak gauge fields. As a matter of fact,

\begin{equation}
Q_{Y}(t) = \int d^{3}x \; J^{o}_{Y} = \frac{N_{\hat{q}}}{6} \;
n_{W}(t) \, ,
\end{equation}

\noindent where $n_{W}$ is the winding number of the gauge
transformation of the pure gauge configuration given by

\begin{equation}
n_{W}(t) = \frac{1}{24 \pi^2} \! \int \!\! d^{3}x \,
\epsilon^{ijk} \, \mbox{tr} (\partial_{i}U U^{-1}
\partial_{j}U U^{-1} \partial_{k}U U^{-1}).
\end{equation}
\\[3pt]
The charge change is

\begin{eqnarray}
\Delta Q_{Y} & = & \displaystyle \frac{N_{\hat{q}}}{6} \left[
n_{W}(t=+\infty) - n_{W}(t=-\infty) \right] \nonumber \\
& = & \frac{N_{\hat{q}}}{6} \; Q_{T} = \frac{N_{\hat{q}}}{6} \;
Q^{(3)} \, n \, ,
\end{eqnarray}

\noindent where

\begin{equation}
Q_{T} = \int d^{4}x \, \partial_{\mu} K^{\mu} = \frac{g^{2}}{16
\pi^{2}} \int d^{4}x \, \mbox{tr} (W_{\mu\nu} \tilde{W}^{\mu\nu})
\, .
\end{equation}
\\[3pt]
A $Z_{3}$ counting number $Q^{(3)}$ is introduced with the
topological charge $n$, equal to $\pm 1$ for nontrivial topology
and $0$ for trivial topology, just as if the topological charge
were itself a $Z_{3}$ charge. It is due to the hypotheses of the
model. A consequence of all these is that each prequark has to
change its hypercharge by the same value, a charge shift which can
be adjusted to cancel anomalies:

\begin{equation}
Y(\hat{q}) \rightarrow Y(\hat{q}) + \frac{n}{3} \;
Q^{(3)}(\hat{q}) = Y(\hat{q}) - \frac{n}{3} \; (B-3L)(\hat{q}) \,
.
\end{equation}
\\[2pt]
The required value for the topological index is $4$, because in
such a case we have

\begin{eqnarray}
\sum_{q_{L} \ell_{L}} Y = 0 \, , \qquad  \displaystyle \sum_{q_{L}
\ell_{L}} Y^{3} - \sum_{q_{R} \ell_{R}} Y^{3} = 0 \, .
\end{eqnarray}

\noindent It is worth noting that the value $n=4$ of the
topological charge does not depend on the relation $Q=T_{3}+Y/2$
between electric charge, weak isospin and hypercharge adopted in
this work, as remarked in Section~\ref{QLsym}.

The charge normalization restores gauge invariance, breaks
presymmetry between prequarks and leptons and dresses prequarks
into fractionally charged topological quarks, consistent with
Eq.~\ref{Ycharges} and our hypotheses concerning the quark-lepton
charge symmetry in Figure~\ref{table1}. It is seen that the
fractional charge of topological quarks is explained as in
condensed matter physics~\cite{EAM1}. Now one can define an
effective current by
\begin{equation}
\displaystyle \hat{J}^{\mu}_{Y,\mbox{\footnotesize eff}} = -
\frac{2}{3} \! \left[ \overline{\hat{q}}_{aL} \gamma^{\mu} (B-3L)
\hat{q}_{aL} + \overline{\hat{q}}_{aR} \gamma^{\mu} (B-3L)
\hat{q}_{aR} \right]
\end{equation}
\\[2pt]
to absorb topological effects: gauge anomaly cancellation, trivial
topology and fractional charge. Thus Eq.~(\ref{newJ}) takes the
gauge-independent form

\begin{eqnarray}
J^{\mu}_{Y} & = & \overline{\hat{q}}_{aL} \gamma^{\mu}
\frac{Y-4(B-3L)/3}{2} \, \hat{q}_{aL} \nonumber \\ & & \nonumber \\
& & + \; \overline{\hat{q}}_{aR} \gamma^{\mu}
\frac{Y-4(B-3L)/3}{2} \, \hat{q}_{aR} \nonumber \\ & & \nonumber \\
& & + \; \overline{\ell}_{aL} \gamma^{\mu} \frac{Y}{2} \,
\ell_{aL} + \overline{\ell}_{aR} \gamma^{\mu} \frac{Y}{2} \,
\ell_{aR} \, .
\end{eqnarray}

\noindent At this point, prequarks can be identified as quarks of
fractional charge and trivial topology: $\hat{q} \rightarrow q$.
Connection between topological quarks and standard quarks is
nonperturbative.

Cancellation of gauge anomalies in the scenario of preleptons of
fractional hypercharge is done in a similar way~\cite{EAM1}. We
just indicate that the topological structure and charge
dequantization in preleptons, which are symmetric to the ones in
topological quarks, are annulled by the charge normalization
procedure leading to leptons with trivial topology and entire
charge, as in prequarks (see Eqs. \ref{Ycharges} and
\ref{Yconstraints}).

\section{Presymmetry beyond the Standard Model}
\begin{figure}[b]
\centering
\includegraphics[width=80mm]{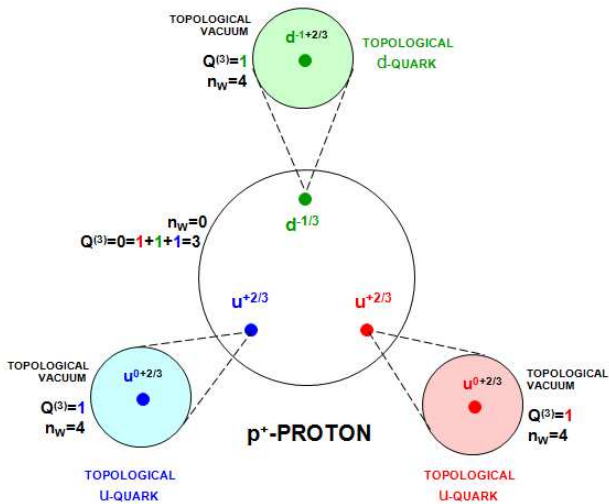}
\caption{Topological quarks in proton.} \label{fig1}
\end{figure}

We here show figures to clarify some points of presymmetry at the
level of the SM. We take the example of proton and its three
quarks in Figure~\ref{fig1}. For each quark, there is a
topological quark. Quarks and topological quarks are different
entities. But three topological quarks turn into three quarks via
a vacuum tunnelling event, a four-instanton~\cite{EAM1}. Quarks
and proton have trivial topology and no weak topological charge.

\begin{figure}[t]
\centering
\includegraphics[width=80mm]{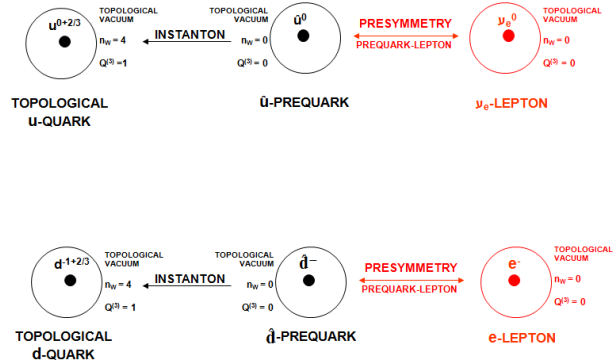}
\caption{Presymmetry between prequarks and leptons.} \label{fig2}
\end{figure}

The connection between topological quarks of fractional charge and
topologically trivial prequarks of integer charges, and the
symmetry between prequarks and leptons, are displayed in
Figure~\ref{fig2}.

Presymmetry is hidden in the SM with no direct implications to be
observed. In fact, it is a pre-dynamical symmetry. Fractional
charge is generated in a peculiar manner but only mathematically.
Physically, there is nothing new at the level of the SM. Nothing
has been altered. This is bad, because one can ask for Occam's
razor: ``Entities should not be multiplied unnecessary.''  It is
the way science is done! To avoid it, a residual presymmetry in
the sense of Ekstein~\cite{Ekstein1,Ekstein2} has to be generated.
This is one of the motivations to going with presymmetry beyond
the SM.

A residual presymmetry requires a doubling of the SM particles.
New families must be nonsequential, duplicating gauge groups.
Other motivations for the duplication of the SM are to extend
presymmetry from matter to forces and extend presymmetry from the
electroweak to the strong sector.

\begin{figure}[b]
\centering
\includegraphics[width=80mm]{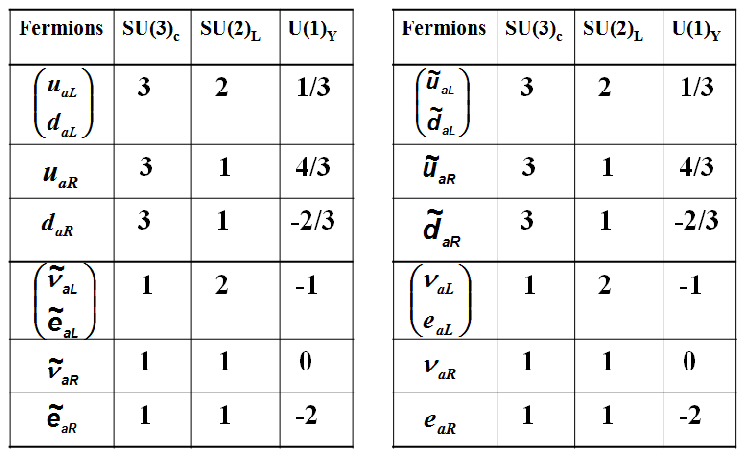}
\caption{Exotic duplication of the SM.} \label{table2}
\end{figure}

Regarding the duplication of the SM, we have been involved in an
exotic version~\cite{EAM3}, where there is an electroweak
separation of quarks and leptons as illustrated in
Figure~\ref{table2}. Quark and lepton partners are denoted by
$\tilde{q}$ and $\tilde{\ell}$, respectively.

\begin{figure}
\centering
\includegraphics[width=80mm]{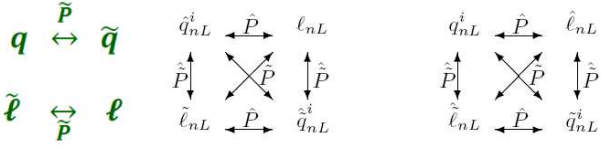}
\caption{Residual presymmetry under duplication of the SM.}
\label{fig3}
\end{figure}

\begin{figure}[t]
\centering
\includegraphics[width=80mm]{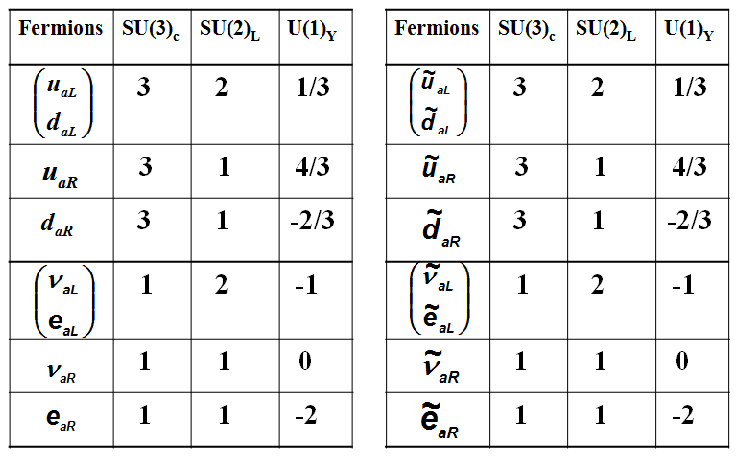}
\caption{Hidden copy of the SM.} \label{table3}
\end{figure}

Under a duplication of the SM, a residual presymmetry is produced
as described in Figure~\ref{fig3} \cite{EAM2}. There are two types
of presymmetries between prequarks and leptons. They introduce an
exotic presymmetry $\tilde{P}$ between prequarks and between
leptons. After charge shifts, presymmetries between prequarks and
leptons are broken. However, exotic symmetry remains exact in
spite of the symmetry-breaking effects. This exotic symmetry can
be interpreted as a manifestation of a residual presymmetry in
accordance with the idea of Ekstein and its extension from matter
to forces. Clearly, particle partners are required to have a
residual presymmetry. Its spontaneous breaking occurs because of
gauge symmetry breaking according to the pattern

\begin{eqnarray}
& \left[ \mbox{SU(3)}_{c} \right]^2 \times \left[ \mbox{SU(2)}_{L}
\right]^2 \times \left[ \mbox{U(1)}_{Y} \right]^2 \times
\tilde{\mbox{P}} & \nonumber \\ & \downarrow & \nonumber \\
& \left[ \mbox{SU(3)}_{c} \right]^2 \times \mbox{SU(2)}_{L} \times
\mbox{U(1)}_{Y} & \nonumber \\ & \downarrow & \nonumber \\ &
\left[ \mbox{SU(3)}_{c} \right]^2 \times \mbox{U(1)}_{em} , &
\end{eqnarray}
\\[3pt]
which involves a duplication of the Higgs sector~\cite{EAM3}.

We are now turning toward a more standard scenario, where there is
no electroweak separation of quarks and leptons. It is the hidden
version, where all standard particles are neutral with respect to
the hidden gauge group. The simplest case is shown in
Figure~\ref{table3}, where the duplication of the SM preserves
spin and handness.

The residual presymmetry is produced in a way similar to that of
the exotic version. Its breaking is also due to the gauge symmetry
breaking, involving duplication of a Higgs sector. This is work in
progress and results will be reported elsewhere.

\section{Conclusions}
Since presymmetry is difficult to be tested at the level of the
SM, it may be discarded by using Occam's razor. But presymmetry is
also difficult to be refuted. Partner particles with a residual
presymmetry are then required to resolve this impasse. The
simplest duplication of the SM keeps spin and handness, extending
the residual presymmetry from matter to forces. It is an up-scaled
copy of the SM particles which appears nondegenerated about the
TeV scale, much as the second and third generations of quarks and
leptons are mere up-scaled copies of the first generation.
Everything for presymmetry! Phenomenological implications are
similar to those of other popular models such as supersymmetric
models with R-parity, little Higgs models with T-parity and
universal extra dimension models with KK-parity, which also
propose duplication of known particles about the TeV scale.

Presymmetry remains hidden and the model is in trouble if there is
no heavy copy of the SM particles. Majorana neutrinos and
sequential families, such as a fourth generation, also bring
problems to the idea of presymmetry which demands that the number
of fermion families and the number of colors be equal. If anything
of this could occur, we would go back to the starting point and
state that the quark-lepton charge symmetry presented in
Figure~\ref{table1} in support of presymmetry is accidental and
not real, which is really hard to be accepted. Hence, our claim is
that a simple doubling of the SM particles which pairs separately
matter and forces should exist, neutrinos should be of Dirac type
and no new sequential family of quarks and leptons should be
found.

Finally, we mention that our work has been mostly theoretical.
More precise phenomenological analyses on the predicted
duplication of the SM particles are necessary. They provide
research opportunities.

\begin{acknowledgments}
This work was supported in part by the Departamento de
Investigaciones Cient\'{\i}ficas y Tecnol\'ogicas, Universidad de
Santiago de Chile, Usach.
\end{acknowledgments}

\bigskip

\end{document}